\documentclass[useAMS,usenatbib]{mn2e}

\usepackage{times}
\usepackage{color}
\usepackage{graphicx}
\newcommand{\kms}{\mbox{$\mathrm{km\,s^{-1}}$}}
\newcommand{\mum}{\mbox{$\mathrm{\mu m}$}}
\newcommand{\Msun}{\mbox{$M_\odot$}}
\newcommand{\Lsun}{\mbox{$L_\odot$}}
\newcommand{\Rsun}{\mbox{$R_\odot$}}
\newcommand{\Mdot}{\mbox{$\dot{M}_{\rm wind}\!=\!1\times10^{-4}\,\Msun\,\mathrm{yr}^{-1}$}}
\newcommand{\gtapprox}{\raisebox{-0.5ex}{$\,\stackrel{>}{\scriptstyle\sim}\,$}}
\newcommand{\ltapprox}{\raisebox{-0.5ex}{$\,\stackrel{<}{\scriptstyle\sim}\,$}}

\title[SS\,433 accretion disc inflow \& outflow]{Inflow {\it and}
  outflow from the accretion disc of the microquasar SS\,433: UKIRT
  spectroscopy}

\author[Perez and Blundell]{Sebastian Perez M.$^{1}$\thanks{E-mail:
    s.perez2@physics.ox.ac.uk } and Katherine
  M. Blundell$^{1}$\\ $^{1}$University of Oxford, Department of
  Physics, Keble Road, Oxford, OX1 3RH, U.K.  }

\begin{document}

\date{}
\pagerange{\pageref{firstpage}--\pageref{lastpage}}
\pubyear{2009}
\maketitle
\label{firstpage}

\begin{abstract}
  
  A succession of near-IR spectroscopic observations, taken nightly
  throughout an entire cycle of SS\,433's orbit, reveal (i) the
  persistent signature of SS\,433's accretion disc, having a rotation
  speed of $\sim\!500$~\kms\, (ii) the presence of the circumbinary
  disc recently discovered at optical wavelengths by Blundell, Bowler
  \& Schmidtobreick (2008) and (iii) a much faster outflow than has
  previously been measured for the disc wind. From these, we find a
  much faster accretion disc wind than has noted before, with a
  terminal velocity of $\sim$1500~\kms. The increased wind terminal
  velocity results in a mass-loss rate of
  $\sim$$10^{-4}$~\Msun~yr$^{-1}$.  These, together with the newly
  (upwardly) determined masses for the components of the SS\,433
  system, result in an accurate diagnosis of the extent to which
  SS\,433 has super-Eddington flows. Our observations imply that the
  size of the companion star is comparable with the semi-minor axis of
  the orbit which is given by $\sqrt{1-e^2}~40~\Rsun$, where $e$ is
  the eccentricity. Our relatively high spectral resolution at these
  near-IR wavelengths has enabled us to deconstruct the different
  components that comprise the Brackett-$\gamma$ line in this binary
  system, and their physical origins. With this line dominated
  throughout our series of observations by the disc wind, and the
  accretion disc itself being only a minority ($\sim$15~per~cent)
  contribution, we caution against use of the unresolved
  Brackett-$\gamma$ line intensity as an ``accretion signature" in
  X-ray binaries or microquasars in any quantitative way.

\end{abstract}

\section{Introduction}\label{sec:intro}

SS\,433, named in 1977 as Stephenson and Sanduleak's 433rd entry in
their catalogue of H$\alpha$ emitters, was first recognised as an
X-ray and radio source by \citet{clark78}, possibly asssociated with
the supernova remnant W50. It became famous as the first known
relativistic jet source in the Galaxy, and is the only one known with
baryonic content in its jets \citep[see comprehensive review
  by][]{fab04}. Its jets are believed to emerge from an accretion
disc, but direct evidence for this has thus far been scant.

The mass of the system has been elusive over the last 30 years. For
example, \citet{hil08}, from observations of absorption lines,
reported masses of $\sim 12$ and $\sim 4$~\Msun\ for the donor star
and the compact object, respectively. On the other hand, \citet{dod91}
had previously suggested that the masses of the components were rather
low, $\sim$3.2~\Msun\ and $\sim 0.8$~\Msun\ for the donor and the
compact object, respectively. \citet{lop06}, looking at the eclipses
seen in X-ray spectra, suggested masses of $\sim 35$ and $\sim
20$~\Msun\ for the donor and the compact object, respectively. Recent
analysis of the stationary H$\alpha$ line has shown, from dynamical
considerations, that the enclosed mass for the whole system is quite
large, around 40~\Msun, and that the mass for the compact object plus
accretion disc attains $16$~\Msun, therefore also implying that the
identity of the compact object is a rather massive stellar black hole
\citep*{kmb08}.

At near-infrared wavelengths, SS\,433 is characterised by a red
continuum and bright emission lines \citep{tho79,mca80}. As at optical
and X-ray wavelengths, these emission lines can be divided into two
groups: (a) {\it stationary} lines, although highly variable in
strength and profile shape; (b) {\it moving} lines, thought to
originate in two oppositely-directed relativistic jets of baryonic
content (moving with speed $v \sim 0.26\,c$). The stationary emission
lines are thought to be produced in the accretion flow and in an
expanding, geometrically thick environment fed from the stellar wind
and the super-Eddington accretion disc outflow \citep{gies02, fab04}.

The system shows three main periodicities: the binary's orbital
motion, with a period of about 13.1 days \citep{cra81}; and the jet
axis's precession and nutation, with periods of about 162 and 6 days,
respectively. A configuration where the jet axis undergoes a
precession cycle in 162 days, referred as the \textit{kinematical
  model}, was proposed to describe the motion of the lines
\citep{fab79,mil79}. In order to account for the 6-day nodding cycle,
\citet{kat82} proposed a dynamical model where the companion exerts a
gravitational torque on the disc periphery.

Near-IR light can escape high opacity and dusty environments more
easily than H$\alpha$ photons can, making it possible to detect
heavily obscured line-emitting regions. We carried out a line
de-blending procedure based on fitting Gaussian profiles to our
near-IR spectra, choosing Brackett-$\gamma$ (hereafter Br$\gamma$) at
$\lambda$\,2.165~\mum\ as the most suitable emission line to model.

The Br$\gamma$ recombination line has been extensively used in the
past to diagnose the presence of accretion discs and outflows
\citep[e.g.,][]{ban97,sha99}. In the case of cataclysmic variables
(CVs), \citet{dhi95} argued that since the Brackett and {He\,\sc i}
emission lines are so strong in emission and also so broad that they
must originate in the accretion disc. In low-mass X-ray binaries
(LMXBs) the Br$\gamma$ line has been detected showing a double-peaked
profile characteristic of accretion discs. \citet{ban97} used the
distance between the peaks to calculate properties of the orbits in
LMXBs, assuming that the compact object accretes via Roche-lobe
overflow.

In this work, we present unprecedentedly high signal-to-noise
mid-resolution near-IR spectra of SS\,433 from UKIRT that has enabled
us to identify the accretion disc and its outflow. In
Section~\ref{sec:obs}, we begin by giving a brief description of the
UKIRT observations and the data reduction. In Section~\ref{sec:brg} we
discuss the results of deconstructing the Br$\gamma$ emission line in
order to identify and quantify the different components present in
emission in the system throughout an orbital period. Finally we
discuss and give concluding remarks on this work in
Sections~\ref{sec:ana} and~\ref{sec:con}.

\section{Observations and data reduction}\label{sec:obs}

We observed SS\,433 every night over an entire orbital cycle with the
{\sc ukirt uist} spectrograph \citep{uist} from 2006 August 17 to
August 29, during which time the precession phase varied from 0.40 to
0.47 respectively. We use the convention in which orbital phase
($\phi_\mathrm{orb}$) zero is when the donor star is eclipsing out the
compact object \citep{gor98}. Precessional phases
($\psi_\mathrm{pre}$) are based on the ephemeris reported in
\citet{fab04}, where precession phase zero is when the jet lines are
maximally separated hence the inclination of the jet axis with our
line-of-sight attains a minimum, i.e., it corresponds to maximum
exposure of the accretion disc to the observer.

{\sc uist} is a 1--5~\mum\ imager-spectrometer with a \mbox{$1024
  \times 1024$} InSb array. The plate scale in spectroscopic mode is
$0 \farcs 12~\mathrm{pixel}^{-1}$. We used a slit of width $0 \farcs
24$. The spectral resolution $R=\lambda/\Delta\lambda$ is about 4000
for the $J$-band and 3900 for the $K$-band. The top of the slit was
pointing to the east and centred on SS\,433's optical position during
acquisition. There were no problems in acquiring the target since
SS\,433 is a very bright red object with $K\sim\!8$~mag. Our data
consist of 13 nights of $K$-band high-resolution spectra and one
night, at primary minimum, covering the complete near-IR window ($J$,
$H$ and $K$-bands).

The observing strategy consisted of taking about forty frames with
exposure times of 45~s each, following a ABBA nodding pattern. The
pixel-to-pixel response of the array was corrected using the
normalised, reduced flat field. Bad pixels, correction of detector
defects and cosmic-ray removal were performed using standard median
filtering techniques. Frequent observations of a standard star
(BS6697) and the arc lamp were performed during each night to provide
an accurate estimate of the wavelength axis. Wavelength calibration
was carried out by fitting a high-order polynomial to the Argon lamp
arc spectra, taken each night. In order to remove the sky
contribution, pairs of nodded frames were subtracted. Standard star
division was applied to each frame in order to flux calibrate the
data. The intrinsic photospheric features were carefully removed from
the standard star by interpolating or fitting a Voigt profile when
needed. Great care was taken to cancel the telluric features,
especially the CO$_2$ absorption near 2.05~\mum. Finally, the spectra
were extracted using Horne's optimal extraction algorithm
\citep{horne} and then combined to produce an average spectrum for
each night with an exposure time of about 2~ks giving a
signal-to-noise $\sim 400$ in the Br$\gamma$ line.

\begin{figure}
  \centering\includegraphics[width=.8\columnwidth]{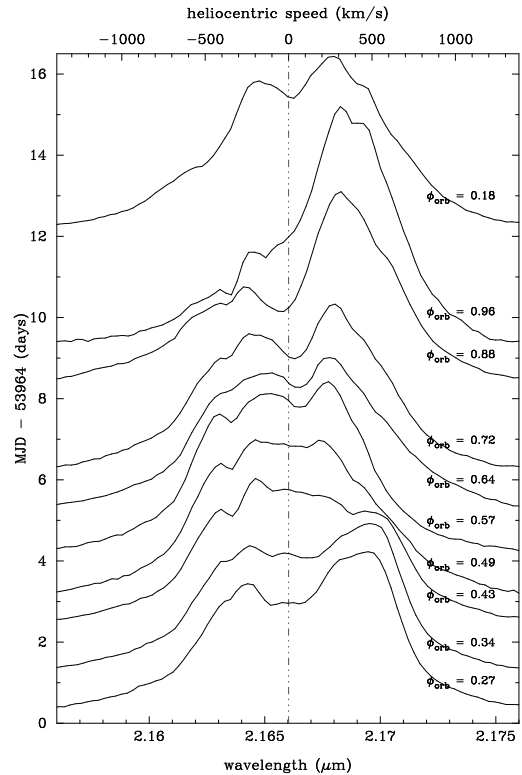}
    \caption{Individual Br$\gamma$ stationary profiles plotted as a
      function of wavelength and modified Julian date of
      observation. Their continua are normalised to unity and aligned
      with the mean day of their observation. Each spectrum is
      labelled with its orbital phase on the right hand side. The
      dashed line corresponds to the systemic velocity measured from
      the {Mg\,\sc ii} radial velocity curve (Perez et al. {\it in
        prep}).}
    \label{fig:brg:evol}
\end{figure}

All data reduction and analysis were carried out using the Perl Data
Language.\footnote{\texttt{http://pdl.perl.org}.}  The reduced spectra
centred on the heliocentric Br$\gamma$ line are displayed in
Fig.~\ref{fig:brg:evol} from which it is immediately apparent that
there is considerable structure in this emission line, and
considerable variation of this structure. We deconstruct these
structures in the next section.

\section{Deconstructing the Brackett-$\gamma$ profile}
\label{sec:brg}

\begin{figure}
  \centering
  \includegraphics[width=\columnwidth]{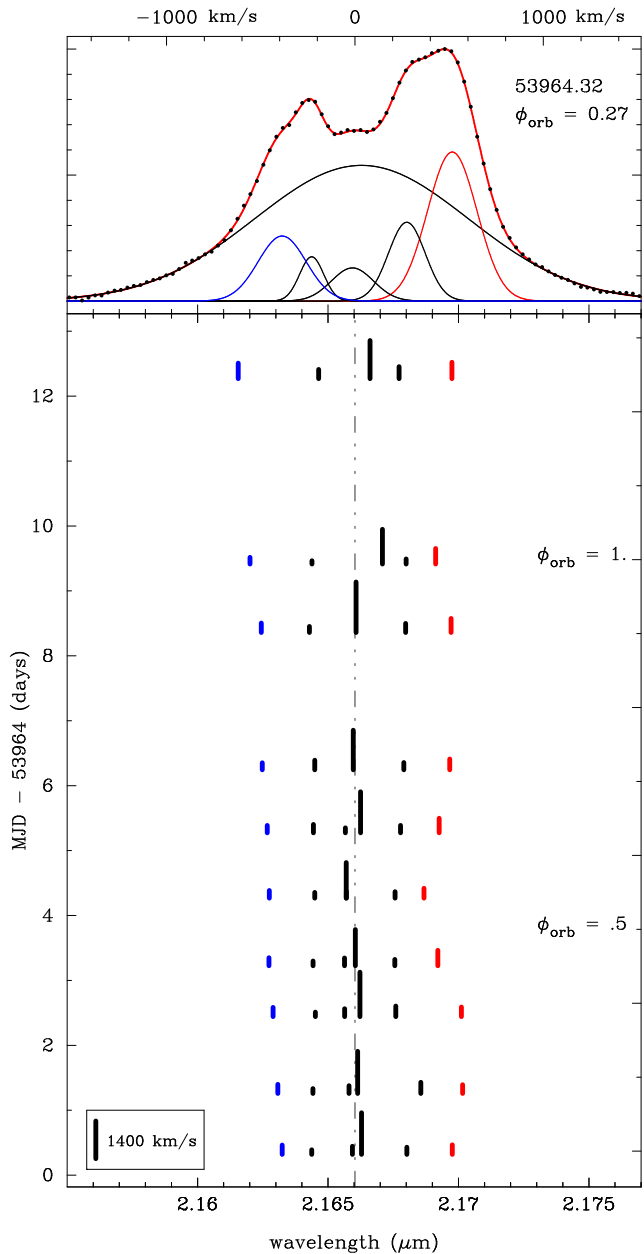}
  \caption{\textit{Upper panel:} Example of Br$\gamma$ stationary
    emission lines observed at orbital phase $\phi_{\rm orb}=0.27$;
    the top $x$-axis corresponds to heliocentric speed in units of
    \kms. \textit{Lower panel:} Tracks of the centroids of the
    Gaussian components fitted to each of our spectra. The modified
    Julian date (MJD) increases vertically and the tick mark heights
    are proportional to the {\sc fwhm} of each component (see inset in
    the bottom left corner). }
  \label{fig:brg}
\end{figure}

Many other authors have studied the stationary H$\alpha$ line profile
in detail \citep[e.g.,][]{fal87, gies02}, albeit over rather shorter
observing runs. Decomposition of a line profile as the sum of Gaussian
components is a technique widely used to extract information from
different parcels of gas in a line emitting region. For example, in
the case of Seyfert galaxies it allows one to distinguish the narrow
line region from the broad line region
\citep[e.g.,][]{ho97}. \citet{kmb08} decomposed the stationary
H$\alpha$ line (observed during a quiescent period in SS\,433's
behaviour) into primarily 3 components: one broad component ({\sc
  fwhm} $\sim\!700$~\kms) whose width was observed to decrease with
precessional phase (i.e., as the jets become more in the plane of the
sky) identified as the accretion disc wind, and two narrower, red- and
blue-shifted components, but stationary in wavelength, being radiated
from a glowing circumbinary ring.

The stationary Br$\gamma$ emission lines show a much more complex
profile than the quiescent H$\alpha$ line studied by
\citet{kmb08}. After trying with different numbers of Gaussian
components we came to the realisation that up to six components were
needed to account for the complexity of the Br$\gamma$ profile
shape. Fig.~\ref{fig:brg} (upper panel) shows an example of a fitted
Br$\gamma$ profile; in this case six Gaussians were used in the fit.

The presence of a P Cygni feature in the stationary lines has been
noted by several authors at certain precessional phases
\citep{cra81,fili88,gies02}. An absorption feature in the blue wing of
the line profile would indeed complicate the analysis and it would
have to be taken into account using models of outflowing winds
\citep[e.g.,][]{cas79}. However, we have found no clear evidence of
the presence of such absorption feature at the epochs at which we
observed SS\,433 (see e.g., top panel of Fig.~\ref{fig:brg}).

\begin{figure}
  \centering\includegraphics[width=.9\columnwidth]{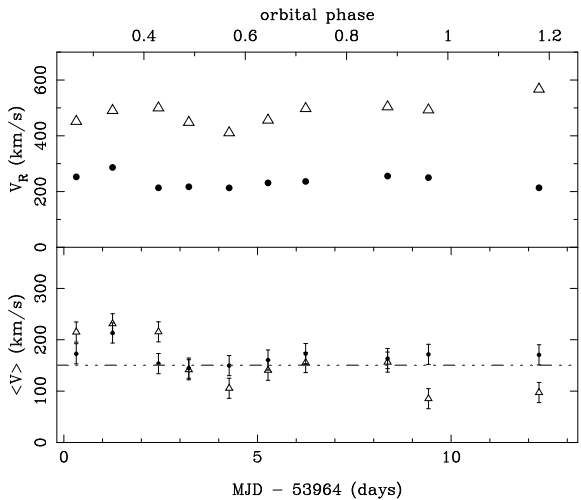}
  \caption{Rotational (upper) and mean (lower) speeds of the rapidly
    rotating components in the Br$\gamma$ line. Open triangles
    correspond to the widely (500~\kms) separated disc lines, while
    solid circles correspond to the circumbinary ring lines. The
    rotational speed is calculated from half the difference of the
    redshifts of the red and blue components in
    Fig~\ref{fig:brg}. Symbol sizes correspond to 1~$\sigma$ error
    bars.}
  \label{fig:rbrg}
\end{figure}

Fig.~\ref{fig:brg} (lower panel) shows the components of the
stationary Br$\gamma$ line as a function of time. It is easy to see
that the Br$\gamma$ complex can be decomposed in three main
constituents: a very broad wind component present at all times in our
data-set and two sets of narrower pairs. The broad wind and both
narrower pairs show a mean velocity close to the systemic speed $V_0
\simeq\!150$~\kms\ (see lower panel in Fig.~\ref{fig:rbrg}). This
measurement of the systemic velocity was obtained from the radial
velocity curve of the {Mg\,\sc ii} $\lambda$ 2.404~\mum\ stationary
line (Perez et al. {\it in prep}). Although this value seems to fit
well with all the stationary emission lines in our near-IR spectra, it
is very different compared to other previous observations (e.g.,
27~\kms\ by \citet{cra81}, 73~\kms\ by \citet{hil08} and $-44$~\kms by
\citet{gies02}).

The broad components plotted in Fig.~\ref{fig:brg} show {\sc fwhm}s
from $1300$ up to $1500$~\kms. The presence of a broad wind component
has been reported before from H$\alpha$ stationary line analysis but
with {\sc fwhm} reaching only up to 800~\kms\ \citep{fal87} and
700~\kms\ during the quiescent period reported by \citet{kmb08},
although broader widths have been observed before and during a flare
by Blundell et al. {\it in prep}.

The inner set of narrow lines, moving at speeds $\sim\!200$~\kms, are
fairly steady in wavelength, in excellent agreement with the presence
of a circumbinary ring reported by \citet{kmb08}.

The most striking discovery that arises from the Br$\gamma$ line
fitting is the presence of a pair of widely-separated, hence rapidly
rotating, narrow components. An example of these narrow (but widely
separated) pairs are depicted in red and blue colours in the upper
panel of Fig.~\ref{fig:brg}. The speed with which the radiating
material spirals in the accretion disc corresponds to half the
difference of the speed of those lines, under the assumption that the
fitted centroids correspond to the tangent speed.  This reveals
material that is spiralling in the potential well at speeds of about
500~\kms\ (see upper panel in Fig.~\ref{fig:rbrg}). The accretion disc
lines also show a somewhat sinusoidal variation, as can be seen in
Fig.~\ref{fig:rbrg}, with a peak-to-peak amplitude of $\sim
100$~\kms.

The sixth component, that only appears between $\phi_\mathrm{orb} =
0.2$ and 0.6, seen in Fig.~\ref{fig:brg} as a small Gaussian blueward
of the position of the broad wind, may correspond to much more
extended line emission that is somehow bonded to the system, hence it
moves at about the systemic velocity but it does not obviously show
any of the characteristic periodicities.

\section{Discussion}\label{sec:ana}

\subsection{The accretion disc and its persistent appearance in the near-IR}\label{subsec:ad}

\begin{figure}
  \centering
  \includegraphics[width=.8\columnwidth]{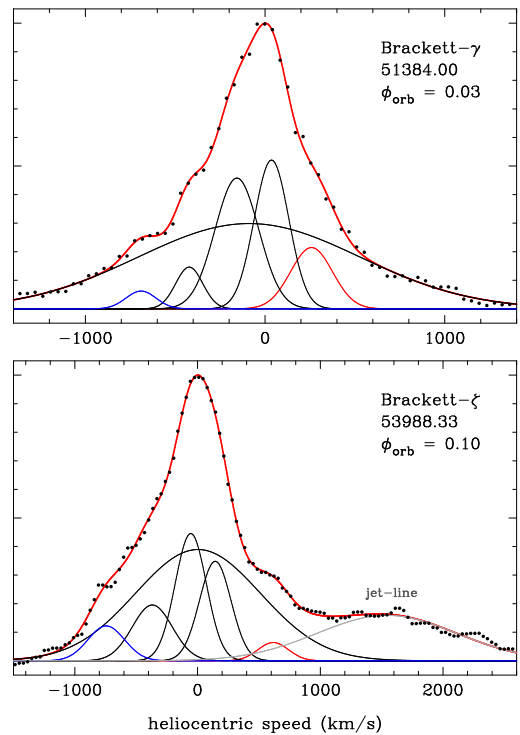}
  \caption{\textit{Upper panel:} Archival CSG4 spectrum of Br$\gamma$
    stationary line taken on 1999 July 25 at orbital phase
    $\phi_{\mathrm{orb}}=0.03$ as a function of heliocentric speed in
    units of \kms. \textit{Lower panel:} Brackett-$\zeta$ spectrum
    taken on the last night of our observing run (2006 September 10)
    at orbital phase $\phi_{\mathrm{orb}}=0.1$ as a function of
    heliocentric speed. Both observations correspond to precessional
    phase $\sim$0.5. The modified Julian day of each observation is
    specified on each plot.}
  \label{fig:check}
\end{figure}

Fleeting glimpses of widely separated pairs of lines centred on
H$\alpha$ have been observed very briefly by \citet{fal87}, and
immediately prior to a flare by Blundell et al. {\it in prep},
implying a high rotation speed. It is remarkable that the lines having
a rotation speed of 500~\kms\ seem to be persistent in near-IR
observations, rather than fleeting or associated with flares as in the
optical: we have no reason to believe that any of our near-IR
observations were during, or just before, a flare.\footnote{We checked
  the radio monitoring records of the Ryle telescope (Guy Pooley) and
  the RATAN telescope (Sergei Trushkin).} Moreover, other near-IR
observations, both our own in $H$-band observations of
Brackett-$\zeta$ 1.74~\mum\ and archival CGS4 observations, are
consistent with the presence of a pair of lines whose positions
correspond to a rotation speed of $\gtapprox 500$~\kms. These are
depicted in Fig.~\ref{fig:check}. For a test particle to move with a
Keplerian velocity of $\sim$500~\kms\ around an object with a mass of
$\sim$16~\Msun\ \citep[as estimated by ][]{kmb08} it would have to be
located at approximately 12~\Rsun\ away from the central object. This
gives us an idea of the extent of the accretion disc.

Several authors have utilised hydrogenic lines to study accreting
systems \citep[e.g., ][]{ban97,sha99}. From observations of low-mass
X-ray binaries, \citet{ban99} have revealed the presence of outflows
and winds in some accreting mass objects. We emphasise the fact that
using the Br$\gamma$ line as a signature of accretion has to be
undertaken with care since the Br$\gamma$ emission is telling us about
outflow via various means, not merely accretion inflow onto a compact
object.

\begin{figure}
  \centering
  \includegraphics[width=\columnwidth]{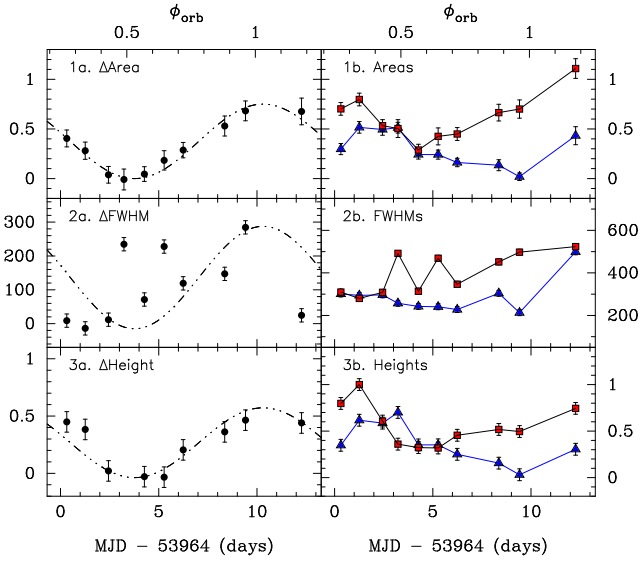}
  \caption{\textbf{Panel 1a:} Difference between the areas of the two
    widely-separated ($500$~\kms) components, red- minus blue-shifted
    (the {\em y}-axis is in units of $10^{-15}$~W~m$^{-2}$). Notice
    the orbital phase dependence (the dashed line is the sinusoidal
    variation corresponding to that orbital phase). \textbf{Panel 1b:}
    Areas of red and blue accretion disc lines as a function of
    time. \textbf{Panels 2a} and \textbf{2b} are the same as in the
    upper panel but for the {\sc fwhm} (in \kms) of each accretion
    disc line. \textbf{Panels 3a} and \textbf{3b} are the equivalent
    plots for the {\sc heights} of the Gaussians fitted to the red and
    blue components, in units of
    $1.6\!\times\!10^{-13}$~W~m$^{-2}$~$\mu$m$^{-1}$. Error bars
    correspond to 1~$\sigma$ uncertainties. }
  \label{fig:areas}
\end{figure}

The area of each accretion disc line (i.e., the components moving at
$\sim\!500$~\kms) seems to be correlated: the difference between the
areas of these two lines shows a very clear orbital phase dependence
(see panel 1a in Fig.~\ref{fig:areas}). As we can see in panel 2b of
Fig.~\ref{fig:areas} the {\sc fwhm} of each line does not show
periodic variations, it is around 300~\kms with a scatter of
$\sim\!70$~\kms. On the other hand, panel 3a of the same figure shows
that the {\sc height} of each line clearly depends on orbital
phase. This behaviour is what we would expect if the material emitting
these components comes from the disc, because as the star, together
with the spherically symmetric wind/corona, transits in front of the
disc it partially obscures the emitting regions.

The crucial points of the orbital phase dependence are as follows: (i)
at orbital phase 0.5, the compact object and its accretion disc are
closer along our line of sight than the companion star is --- at this
orbital phase there is no asymmetric obscuration so there is zero
difference in the areas of the red and blue lines.  (ii) At all other
orbital phases, the red line area (and indeed height) exceeds that of
the blue line. From this we draw two conclusions: first there is low
level obscuration from the stellar wind at all orbital phases except
when the disc acts as its own windbreak at orbital phase 0.5. (We note
that most of the time SS\,433's accretion disc is completely obscured
in the optical.)  At all other phases besides 0.5 there is a component
of the stellar wind between us and the disc lines that attenuates the
blue lines somewhat more than the red lines. Second, we conclude that
the sinusoidal variation in area difference does indeed tell us about
the radial extent of the surface of the star, or of its obscuring
stellar wind.  The maximal difference in area occurs at orbital phase
1.0 when the star is closer along our line of sight than the accretion
disc and compact object: this is the phase when the greatest
attenuating path length though the star (or its dense stellar wind) is
presented from the accretion disc along the line of sight to Earth.
The fact that the variation is sinusoidal is consistent with a
reasonably circularly (spherically) symmetric stellar wind.

\subsubsection{Implications for the geometry of the system}
\label{sec:geo}

In the previous sub-section we found that the companion star (and its
wind) eclipses out some areas of the accretion disc. If the orbit of
the system is essentially circular then the variation of the
difference of the accretion disc's lines (Fig.~\ref{fig:areas})
implies that the star has a diameter comparable with the radius of the
orbit. On the other hand, if the orbit is rather elliptical, then the
star is smaller and the semi-major axis of the orbit must be along our
line-of-sight as inferred from the shape of the light curve, namely,
the fact that the secondary minima occurs precisely close to mid-way
between the primary minima \citep[see eg., ][]{kem86,gor98}.

Recently, \citet{kmb08} found that the total mass of SS\,433 is
approximately 40~\Msun. We can roughly estimate the size of the system
to be about twice the semi-major axis of the orbit $a$, as given by
Kepler's third law: $a^3 = G M_\mathrm{sys} (P_\mathrm{orb}/2\pi)^2 $,
where $P_\mathrm{orb}$ is the orbital period and $M_\mathrm{sys}$ is
the total mass of the binary. This relation implies a size for the
whole system of $2a \simeq 160$~\Rsun.  The radius of the companion
star should be $R \sim a/2$. Thus, the eclipsing of the
widely-separated accretion disc lines implies a maximum radius for the
companion star $R \simeq 40$~\Rsun. For other recent mass estimates we
obtain similarly large values for the size of the star: \citet{hil08}
reported $M_\mathrm{sys} \approx 16$~\Msun\ implying $R \sim
29$~\Rsun\ and \citet{lop06} estimated $M_\mathrm{sys} \approx
55$~\Msun\ which yields $R \sim 44$~\Rsun. All of these mass estimates
assume the case where the orbit is essentially circular. This size is
smaller but still comparable with the radius of evolved stars
undergoing severe mass-loss such as $\eta$~Car ($\sim 80~\Rsun$). This
large size is reduced only at the expense of accepting an eccentric
orbit. The size of the companion star in the eccentric case would be
$b=\sqrt{1-e^2}\,40~\Rsun$ where $b$ is the semi-minor axis of the
orbit and $e$ is the eccentricity.

\begin{figure*}
  \centering
  \includegraphics[width=.7\textwidth]{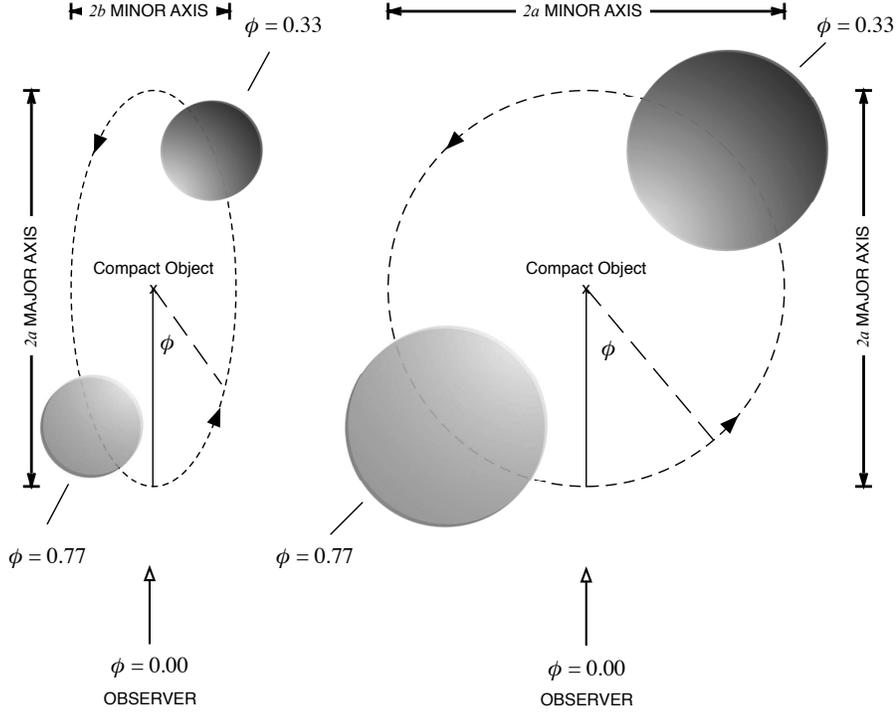}
  \caption{ Diagram that shows the 2 possible extreme scenarios
    concerning the configuration of the SS\,433 binary system: (left)
    eccentric orbit, (right) circular orbit. The semi-major axis of
    the orbit is $a \simeq 80$~\Rsun (see Section~\ref{sec:geo}).}
  \label{fig:orb}
\end{figure*}

\subsection{The disc wind: near-IR view and mass-loss rate}

\subsubsection{Precessional- and orbital-phase dependence, or lack thereof}

Fig.~\ref{fig:brg} shows that the broad wind component, inferred
\citep[on the basis of optical observations,][]{kmb08} to originate
from the disc, lacks indications of orbital and nutational
periodicities manifested by the accretion disc. We have not been able
to test for precessional phase dependence due to our observations
spanning only 0.07 of a precessional phase. A possible explanation of
this could be that the near-IR wind comes from very close into the
centre of the disc, where the temperature of the gas could reach
$\sim\!10^8$~K. At these small radii the material only experiences the
influence of the compact object and any tidal torque from the
companion star could be a negligible perturbation. At greater radii as
observable in the optical, the torques exerted by the companion are
much stronger, hence precessional periodicities are manifested.

\subsubsection{Mass-loss via line-driven winds}

Close binary systems undergo severe mass loss and mass exchange during
their lifetime. Like early-type stars, their spectra show a number of
emission lines due to the large amount of ionised gas present in the
system. This ionised gas produces thermal radio emission and an excess
of continuum photons at infra-red wavelengths, which somewhat
correlates with line emission, yielding similar mass loss
estimates. These three spectral features yield mass-loss rate
estimates for emission line stars \citep{kog07}.

\citet{lei88} has used the luminosity of the optically-thin H$\alpha$
line as a mass-loss tracer. The relation between $\dot{M}$ and
luminosity relies on the dependence of both quantities on the density
of the wind, $\rho(r)$. Following Leitherer's approach we can derive
an equivalent expression for Br$\gamma$ emission, by assuming that the
emission is optically thin in an isothermal wind. Therefore, the
luminosity of a recombination line, $L_l$, will be given by:
\begin{equation} 
  L_l = C_n \, f(T_\mathrm{eff}) \, I(v_0/v_\infty, \beta) \,
  \frac{\dot{M}^2}{\mu^2 v^2_\infty R},
  \label{eq:L_l}
\end{equation}

\noindent where $v_\infty$ is the terminal speed reached by the wind,
$\mu$ is the mean atomic weight and $R$ is the radius of the source
(i.e., where the wind starts), $\dot{M}$ is the mass-loss rate, $I$ is
an integral defined below and $\beta$ is a parameter that controls the
shape of the velocity law of the wind. $C_n$ and $f(T_\mathrm{eff})$
are constants for each transition given by:
\begin{equation}
  C_n = \frac{n_1^2 A_{12} h\nu_{12}}{4 \pi m_\mathrm{H}^2 } \left(
  \frac{h^2}{2 \pi m_\mathrm{e} k} \right)^{3/2},
  \label{eq:C_n}
\end{equation}

\noindent and,
\begin{equation}
  f(T_{\rm eff}) = b_n \, T_\mathrm{e}^{-3/2} \, e^{\chi_n/kT_{\rm
  e}},
  \label{eq:f_T}
\end{equation}

\noindent where $n_1$ and $n_2$ are the principal quantum numbers
corresponding to the upper and lower levels in the atom,
respectively. The parameter $b_n$ denotes the departure coefficient
and it accounts for non-LTE effects (assumed to be $\sim\!1$ for
simplicity). $A_{12}$ is the transition probability, $k$ is the
Boltzmann constant and $\chi_n$ is the ionisation potential of the
level $n_1$. We assume that the wind regions are in radiative
equilibrium, i.e., the system's effective temperature $T_{\rm eff}$ is
very close to the electron temperature, $T_{\rm e}$.

The quantity $I = I(v_0/v_\infty, \beta)$ in equation~\ref{eq:L_l}
represents the integral over the ratio between the dilution factor and
the square of the velocity law $v(r)$. In the case where the velocity
law corresponds to a $\beta$-law proposed by \citet{cas79} and the
dilution factor is the standard formula given by \citet{kog07}, the
integral is given by:
\begin{equation}
  I = \frac{1}{2} \, \int_0^1 \! \frac{1 - \sqrt{1-x^2}}{\left[ u +
  (1-u)(1-x)^\beta \right]^2} \, dx,
  \label{eq:I_r}
\end{equation}

\noindent where $x=R/r$, $r$ being the distance from the centre of the
source, and $u = v_0/v_\infty$. The speed $v_0$ is the initial
velocity of the wind at $r=R$.

\subsubsection{Mass-loss rate via the disc wind}

During our observations, the wind Br$\gamma$ luminosity attains a
maximum at $\phi_{\rm orb} = 0.5$, supporting the idea that this
emission comes from the accretion disc; we have measured this flux
density to be $F({\rm Br}\gamma) = (6 \pm 0.5)\! \times \!
10^{-15}$~W\,m$^{-2}$. The distance to SS\,433 has been accurately
determined by \citet{kmb04} as $d=5.5 \pm 0.2$~kpc. This yields a
luminosity for the wind in Br$\gamma$ of
$14\,\pm\,1$~\Lsun\ (corrected for $A_\mathrm{v} = 7.8$ magnitudes of
optical extinction). We expect the wind to start at least at a radius
$R=20$~\Rsun\ from the compact object, according to the dimensions of
the system (see Fig.~\ref{fig:orb}). We computed the integral in
equation~\ref{eq:I_r} assuming an exponent $\beta \approx 0.7$ and an
initial velocity of the order of the sound speed $v_0 = 15~\kms$ (or
$0.01~v_\infty$), which is a good approximation for hot early-type
stars \citep{lei88}. Because of severe dust extinction towards SS\,433
there are no measurements of the terminal velocity from UV P-Cygni
profiles. Therefore we use the width of the wind component {\sc fwhm}
$\simeq \!  1500~\kms$ as the terminal velocity of the wind. This is a
very modest value compared with some early-type stars
\citep{lei88}. The effective temperature has been estimated to be
$T_{\rm eff} = 5\!\times\!10^4$~K by fitting the optical-UV spectral
flux distribution \citep{gies02}. The mean atomic weight of evolved
massive stars is $\mu \approx 2$ \citep{fuch06}.

If we apply equation~\ref{eq:L_l} to the Br$\gamma$ transition of
hydrogen (i.e., $n_1=7$ and $n_2=4$) we find a useful formula that
relates the mass loss with the luminosity of the wind:

  \begin{equation}
    \dot{M}_{\rm wind} \simeq 1\!\times\!10^{-4}~\Msun\,\mathrm{yr}^{-1}
    \times \mu_2\, v_{1500}\, R_{20}^{1/2} L_{10}^{1/2},
  \end{equation}

\noindent where $\mu_2 = \mu/2$, $v_{1500} = v_\infty/1500~\kms$,
$R_{20} = R/20~\Rsun$ and $L_{10} = L(\mathrm{Br}\gamma)/10~\Lsun$.

This value is the same order of magnitude as the mass-loss rate
estimated by \citet{fuch06} who fitted the mid-IR continuum.  These
estimates, although discrepant by a factor of 2--3, are in resonably
good agreement given that both approaches may suffer from the
assumptions made regarding the geometry of the system and the choice
of some physical parameters.  A crucial unknown is the mean atomic
weight of the wind, since the mass-loss ratio scales with $\mu$ as
$\dot{M} \propto \mu$. \citet{lei95} studied the composition of
early-type stars and from the H/He ratios they inferred values for
$\mu$ between 1.5 and 2.7. Another important source of uncertainty is
the choice of velocity law. For $\beta<1$ the mass-loss rate scales at
most by a factor of 2 and it is quite insensitive to the choice of
$v_0$. On the other hand, for values of $\beta$ greater than one,
$\dot{M}$ depends heavily upon $v_0$. Fortunately, $\beta \simeq\!
0.7-0.8$ for winds in O-stars which may match SS\,433's hot
environment better than B or colder stars (which have $\beta>1$). The
final important effect that must be taken into account is the
clumpiness of the wind. \citet{len04} show that the Br$\gamma$ line
may be especially affected by inhomogeneities in the wind. An
unclumped wind density can be taken as the real wind density
multiplied by a \textit{clumpiness} factor $f$. Thus, when clumpiness
is present in the wind $\dot{M}$ is reduced by a factor $\sqrt{f}$
\citep{len04}. However, the degree of clumpiness is not very high
closer to the source where the Br$\gamma$ line is formed, thus it
should not dramatically affect our estimates.

Although continuum driven winds might contribute to SS\,433's mass
loss \citep{clark07}, they are likely to be of less importance than
line-driven winds in this particular case, because of the presence of
high velocity outflows ($>1400$~\kms\, see Section~\ref{sec:brg}) with
high temperatures, diagnostics which rule out continuum driven winds
as main responsible of mass loss.

\subsection{The circumbinary ring: near-IR view and mass-loss rate}

The idea of the presence of a circumbinary disc in SS\,433 was
considered by \citet{fab93} and recently detected by \citet{kmb08} by
decomposing the H$\alpha$ line profile in three components (during its
quiescent state). In the optical this circumbinary ring of material
orbits the system at a speed of $\sim\!200$~\kms. It has been proposed
that the origin of this glowing material corresponds to overflow of
gas from the L2 point, assuming that the system has filled its Roche
lobe \citep{fab93,kmb08,fili88}. Our Br$\gamma$ decomposition is in
excellent agreement with the presence of this excretion disc since we
see a pair of lines, one permanently blueshifted and the other
permanently redshifted at fairly stable wavelengths, lacking any
orbital phase dependence and rotating at speeds close to $200$~\kms.

\subsection{Mass-transfer rates in SS\,433 and comparisons with the Eddington rate}

According to the basic theory of magneto-hydrodynamic winds from
accretion discs, the mass-loss in the wind is about an order of
magnitude smaller than the mass accretion rate (inflow at the centre
of the disc), i.e., $\dot{M}_{\rm acc} \simeq 10 \,\dot{M}_{\rm wind}$
\citep{kon00}. Based on our estimate of the mass outflow in the wind
\Mdot, this implies an accretion rate of the order of $\dot{M}_{\rm
  acc} \approx \!  10^{-3}~\Msun\,\mathrm{yr}^{-1}$. On the other
hand, the maximum accretion rate allowed by the Eddington criterion,
which assumes spherical symmetry, is $\dot{M}_{\rm acc} = 4 \pi c
m_{\rm p} R_{\rm lso} / \sigma_T$, where $R_{\rm lso} = 6 G M / c^2$
is the radius of the last stable orbit. This yields a maximum
accretion rate of $\dot{M}_{\rm acc}\!  \simeq\!
10^{-7}~\Msun\,\mathrm{yr}^{-1}$ for a black-hole of 16~\Msun. This is
greatly exceeded by our estimate of the accretion rate in the disc,
hence confirming the super-Eddington nature of SS\,433, which is
permitted of course because of the non-spherical geometry of the
accretion.

\section{Concluding remarks}\label{sec:con}

A succession of near-IR spectroscopic observations, taken nightly
throughout an entire cycle of SS\,433's orbit with {\sc uist} on
UKIRT, has revealed:

\begin{description}

\item[(i)] The persistent signature of SS\,433's accretion disc as two
  emission components, blue- and red-shifted, having a rotation speed
  of $\sim\!500$~\kms. The difference in area and in height between
  each component shows a clear orbital phase dependence, very likely
  the signature of the disc being eclipsed out as the star transits in
  front of it. This contributes only about $15\pm5$ per~cent of the
  total Br$\gamma$ emission line in the near-IR.

\item[(ii)] The presence of the circumbinary disc (excretion disc)
  recently discovered at optical wavelengths by \citet*{kmb08}. It
  contributes about $15\pm5$ per~cent of the total emission in
  Br$\gamma$.

\item[(iii)] A much faster outflow than has previously been measured
  from the wind. What supports the model that the wind comes from the
  accretion disc is that the luminosity of the broad component attains
  its maximum at $\phi_\mathrm{orb}\sim 0.5$, when the compact object
  is closer to us (see Section~\ref{subsec:ad}). The presence of this
  faster outflow ($v \sim 1500$ \kms) yields a new upper limit for the
  mass-loss of \Mdot. This outflow corresponds to a very significant
  $70\pm5$ per~cent of the total Br$\gamma$ line emission.

\end{description}

These, together with the newly (upwardly) determined masses for the
components of the SS\,433 system, result in an accurate diagnosis of
the extent to which SS\,433 has super-Eddington flows: its accretion
rate is 10$^4$ times higher than the Eddington limit. Our relatively
high spectral resolution at these near-IR wavelengths has enabled us
to deconstruct the different components, and their physical origins,
that comprise the Brackett-$\gamma$ line in this binary system.  With
this line dominating throughout our series of observations by the disc
wind, and the accretion disc being only a minority
(\ltapprox\,15~per~cent) contribution, we caution against use of the
unresolved Brackett-$\gamma$ line intensity as an ``accretion
signature" in X-ray binaries or microquasars in any quantitative way.

\section*{Acknowledgments}
We are very grateful to the Sciences and Technology Facilities Council
({\sc stfc}) studentship for the support of this research. We warmly
thank the referee for useful comments on the manuscript. The United
Kingdom Infrared Telescope is operated by the Joint Astronomy Centre
on behalf of {\sc stfc} of the U.K. We are especially grateful to the
staff of UKIRT especially Paul Hirst for their flexibility and
assistance they gave us in the accommodation and execution of our
time-resolved observations. K. B. thanks the Royal Society for a
University Research Fellowship.

\bsp

\label{lastpage}


\begin{thebibliography}{99}


\bibitem[Bandyopadhyay et al.(1997)]{ban97} Bandyopadhyay, R.,
Shahbaz, T., Charles, P.~A., van Kerkwijk, M.~H., \& Naylor, T.\ 1997,
MNRAS, 285, 718 

\bibitem[Bandyopadhyay et al.(1999)]{ban99} Bandyopadhyay, R.~M.,
Shahbaz, T., Charles, P.~A., \& Naylor, T.\ 1999, MNRAS, 306, 417


\bibitem[Blundell \& Bowler(2004)]{kmb04} Blundell, K.~M., \& Bowler,
M.~G.\ 2004, ApJL, 616, L159 

\bibitem[Blundell et al.(2008)]{kmb08} Blundell, K.~M., Bowler, M.~G.,
\& Schmidtobreick, L.\ 2008, ApJ, 678, L47 

\bibitem[Castor \& Lamers(1979)]{cas79} Castor, J.~I., \& Lamers,
H.~J.~G.~L.~M.\ 1979, ApJS, 39, 481 

\bibitem[Clark \& Murdin(1978)]{clark78} Clark, D.~H., \& Murdin, P.\
1978, Nature, 276, 44 

\bibitem[Clark et al.(2007)]{clark07} Clark, J.~S., Barnes, A.~D., \&
Charles, P.~A.\ 2007, MNRAS, 380, 263 

\bibitem[Crampton \& Hutchings(1981)]{cra81} Crampton, D., \&
Hutchings, J.~B.\ 1981, ApJ, 251, 604 

\bibitem[Dhillon \& Marsh(1995)]{dhi95} Dhillon, V.~S., \& Marsh,
T.~R.\ 1995, MNRAS, 275, 89 

\bibitem[D'Odorico et al.(1991)]{dod91} D'Odorico, S., Oosterloo, T.,
  Zwitter, T., \& Calvani, M.\ 1991, Nature, 353, 329

\bibitem[Fabian \& Rees(1979)]{fab79} Fabian, A.~C., \& Rees,
  M.~J.\ 1979, MNRAS, 187, 13P 

\bibitem[Fabrika(1993)]{fab93} Fabrika, S.~N.\ 1993, MNRAS, 261, 241

\bibitem[Fabrika(2004)]{fab04} Fabrika, S.\ 2004, Astrophysics and
Space Physics Reviews, 12, 1 

\bibitem[Falomo et al.(1987)]{fal87} Falomo, R., Boksenberg, A.,
Tanzi, E.~G., Tarenghi, M., \& Treves, A.\ 1987, MNRAS, 224, 323

\bibitem[Filippenko et al.(1988)]{fili88} Filippenko, A.~V., Romani,
R.~W., Sargent, W.~L.~W., \& Blandford, R.~D.\ 1988, AJ, 96, 242

\bibitem[Fuchs et al.(2006)]{fuch06} Fuchs, Y., Koch Miramond, L., \&
{\'A}brah{\'a}m, P.\ 2006, A\&A, 445, 1041 

\bibitem[Gies et al.(2002)]{gies02} Gies, D.~R., Huang, W., \&
McSwain, M.~V.\ 2002, ApJ, 578, L67 

\bibitem[Goranskii et al.(1998)]{gor98} Goranskii, V.~P., Esipov,
V.~F., \& Cherepashchuk, A.~M.\ 1998, Astronomy Reports, 42, 209

\bibitem[Hillwig \& Gies(2008)]{hil08} Hillwig, T.~C., \& Gies, D.~R.\
2008, ApJL, 676, L37 

\bibitem[Ho et al.(1997)]{ho97} Ho, L.~C., Filippenko, A.~V., Sargent,
W.~L.~W., \& Peng, C.~Y.\ 1997, ApJS, 112, 391 

\bibitem[Horne(1986)]{horne} Horne, K.\ 1986, PASP, 98, 609

\bibitem[Katz et al.(1982)]{kat82} Katz, J.~I., Anderson, S.~F.,
  Grandi, S.~A., \& Margon, B.\ 1982, ApJ, 260, 780

\bibitem[Kemp et al.(1986)]{kem86} Kemp, J.~C., et al.\ 1986, ApJ,
  305, 805 

\bibitem[Kogure \& Leung(2007)]{kog07} Kogure, T., \& Leung,
  K.-C.\ 2007, The Astrophysics of Emission-Line Stars, by T.~Kogure
  and K.-C.~Leung.~Berlin: Springer, 2007.~ ISBN: 978-0-387-34500-0,

\bibitem[Konigl \& Pudritz(2000)]{kon00} Konigl, A., \& Pudritz,
R.~E.\ 2000, Protostars and Planets IV, 759 

\bibitem[Leitherer(1988)]{lei88} Leitherer, C.\ 1988, ApJ, 326, 356

\bibitem[Leitherer et al.(1995)]{lei95} Leitherer, C., Chapman, J.~M.,
\& Koribalski, B.\ 1995, ApJ, 450, 289 

\bibitem[Lenorzer et al.(2004)]{len04} Lenorzer, A., Mokiem, M.~R., de
Koter, A., \& Puls, J.\ 2004, A\&A, 422, 275 

\bibitem[Lopez et al.(2006)]{lop06} Lopez, L.~A., Marshall, H.~L.,
Canizares, C.~R., Schulz, N.~S., \& Kane, J.~F.\ 2006, ApJ, 650, 338

\bibitem[McAlary \& McLaren(1980)]{mca80} McAlary, C.~W., \& McLaren,
R.~A.\ 1980, ApJ, 240, 853 

\bibitem[Milgrom(1979)]{mil79} Milgrom, M.\ 1979, A\&A, 76, L3

\bibitem[Ramsay Howat et al.(2004)]{uist} Ramsay Howat, S.~K., et al.\
2004, Proc.~SPIE, 5492, 1160 

\bibitem[Shahbaz et al.(1999)]{sha99} Shahbaz, T., Bandyopadhyay,
R.~M., \& Charles, P.~A.\ 1999, A\&A, 346, 82 

\bibitem[Thompson et al.(1979)]{tho79} Thompson, R.~I., Rieke, G.~H.,
Tokunaga, A.~T., \& Lebofsky, M.~J.\ 1979, ApJ, 234, L135

\end{thebibliography}
\end{document}